\documentclass[journal,9pt]{IEEEtran}

\usepackage{amsmath,amsfonts}
\usepackage{algorithmic}
\usepackage{algorithm}
\usepackage{array}
\usepackage{amsmath}
\usepackage[caption=false,font=footnotesize,labelfont=rm,textfont=rm]{subfig}

\usepackage{textcomp}
\usepackage{stfloats}
\usepackage{url}
\usepackage{verbatim}
\usepackage{graphicx}
\usepackage{cite}
\usepackage{hyperref}
\usepackage{cite}
\usepackage{subfig}

\begin{document}

\title{\fontsize{16pt}{19pt}\selectfont High-Resolution Multipath Angle Estimation Based on Power-Angle-Delay Profile for Directional Scanning Sounding}

\author{{Huixin Xu, Jianhua Zhang,~\IEEEmembership{Senior Member,~IEEE,} Pan Tang,~\IEEEmembership{Member,~IEEE,} Hongbo Xing, Lei Tian, and Qixing Wang}
\thanks{This work was supported in part by the National Key R\&D Program of
China under Grant 2023YFB2904805, in part by the National Natural Science
Foundation of China under Grant 62201086, in part by the Beijing Natural
Science Foundation under Grant L243002, and in part by the Beijing University of Posts and Telecommunications-China Mobile Research Institute Joint Innovation Center. \textit{(Corresponding author: Jianhua Zhang.)}}
\thanks{Huixin Xu, Jianhua Zhang, Pan Tang, Hongbo Xing and Lei Tian are with the State Key Lab of Networking and Switching Technology, Beijing University of Posts and Telecommunications, China. (e-mail: xuhuixin@bupt.edu.cn; jhzhang@bupt.edu.cn; tangpan27@bupt.edu.cn; hbxing@bupt.edu.cn; tianlbupt@bupt.edu.cn)}
\thanks{Qixing Wang is with the China Mobile Research Institute, China. (e-mail: wangqixing@chinamobile.com)}
}

\markboth{IEEE Transactions on Antennas and Propagation,~Vol.~8, No. 23, 2023}%
{Shell \MakeLowercase{\textit{et al.}}: A Sample Article Using IEEEtran.cls for IEEE Journals}

\maketitle
\begin{abstract}
Directional scanning sounding (DSS) has become widely adopted for high-frequency channel measurements because it effectively compensates for severe path loss. However, the resolution of existing multipath component (MPC) angle estimation methods is constrained by the DSS angle sampling interval. Therefore, this communication proposes a high-resolution MPC angle estimation method based on power-angle-delay profile (PADP) for DSS. By exploiting the mapping relationship between the power difference of adjacent angles in the PADP and MPC offset angle, the resolution of MPC angle estimation is refined, significantly enhancing the accuracy of MPC angle and amplitude estimation without increasing measurement complexity. Numerical simulation results demonstrate that the proposed method reduces the mean squared estimation errors of angle and amplitude by one order of magnitude compared to traditional omnidirectional synthesis methods. Furthermore, the estimation errors approach the Cramér-Rao Lower Bounds (CRLBs) derived for wideband DSS, thereby validating its superior performance in MPC angle and amplitude estimation. Finally, experiments conducted in an indoor scenario at 37.5 GHz validate the excellent performance of the proposed method in practical applications.

\end{abstract}
 
\begin{IEEEkeywords}
Directional scanning sounding, multipath angle estimation, Cramér–Rao lower bound, channel measurements.
\end{IEEEkeywords}

\section{Introduction}
\IEEEPARstart{T}{he} millimeter-wave (mmWave) and terahertz (THz) bands are promising for achieving higher data transmission rates in future wireless communication systems, owing to their abundant spectrum resources \cite{C1_6G_survey,3D_MIMO1}. Understanding the channel propagation characteristics of the mmWave or THz bands is crucial for the design, development, and optimization of these systems \cite{IoTJ_JT}, and this requires extensive, reliable propagation channel measurement data. The directional scanning sounding (DSS) method, which involves rotating a high-gain horn antenna on a mechanical turntable to probe omnidirectional channels, is widely used in high-frequency channel measurements \cite{C2_Omni_M1_2,C2_Omni_M2_2,DSS_SAGE,DSS_ZRN1}. This is because high-gain antennas can compensate for the significant path loss in the mmWave and THz bands, while also being cost-effective, simple to operate, and easy to implement. 

The existing multipath component (MPC) angle estimation methods for DSS can be divided into three categories. The first category uses high-resolution parameter estimation algorithms, such as space-alternating generalized expectation-maximization (SAGE) \cite{SAGE,C2_JHZ_JSAC,DSS_SAGE}. This method offers high accuracy in estimating MPC parameters. However, it is highly sensitive to phase jitter and faces challenges such as difficulty in obtaining the phase pattern and determining the phase center of the antenna, especially in the THz band. Additionally, it suffers from high computational complexity. The second category treats MPC angle estimation as a spatial deconvolution process of antenna de-embedding. The methods in \cite{DSS_ZRN1} and \cite{DSS_ZRN2} solve this problem using pseudo-inversion and Tikhonov regularization, respectively. However, this method requires the angular sampling interval (ASI) be smaller than the antenna's  half-power beamwidth (HPBW), with ASI set to 2$^{\circ}$ or less, which substantially extends the required measurement time. The third category synthesizes the directional channels of the DSS into an omnidirectional channel \cite{C2_Omni_M1_2,C2_Omni_M2_2}, treating each local maximum (peaks) in synthetic omnidirectional power delay profile (PDP) as an MPC. However, as the directional antenna rotates with a fixed ASI, the angle estimation error can be as large as half the ASI. Furthermore, there is also some error in the power estimation of the MPCs. The MPC angle estimation methods discussed above either require precise antenna responses or consume significant measurement time, or they suffer from insufficient estimation accuracy, failing to strike a balance between measurement complexity and estimation precision.

This communication proposes a high-resolution MPC angle estimation method for DSS (HAED) to address the insufficient accuracy in the third category of methods, significantly improving angle estimation precision without increasing measurement complexity. Specifically, the MPC power difference of adjacent directions based on the power-angle-delay profile (PADP) is calculated, and the offset angle that most likely matches it is found from the known power differences of adjacent ARPs, which subsequently corrects the coarse angle and power estimates of the traditional method. Compared to traditional methods, the proposed approach overcomes the limitation that MPC angle estimation accuracy depends on ASI and avoids issues such as antenna phase pattern measurements and increased measurement time, making it highly valuable in practical applications. The main contributions are as follows:

\begin{itemize}
    \item A high-resolution MPC angle estimation method based on PADP for DSS is proposed, which significantly improves the estimation accuracy of MPC parameters without increasing measurement complexity.
    \item Numerical simulation results show that the proposed method significantly reduces the mean squared estimation error of MPC angle and amplitude, approaching the Cramér-Rao lower bounds (CRLBs) derived for the wideband DSS.
    \item Realistic channel measurements at 37.5 GHz in an indoor scenario validate the accuracy of the proposed method, with angle variations within 1° being accurately estimated.
\end{itemize}

\section{Signal Model and Proposed Method}
\label{sec_2}

\subsection{Signal Model}

\graphicspath{{picture/}}
\begin{figure}[h] 
    \centering  
    \includegraphics[width=8.5cm]{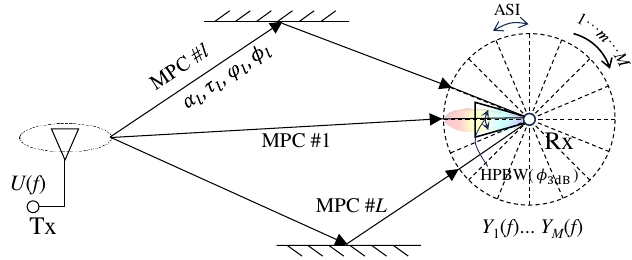}
    \caption{A diagram of the channel measurement based on DSS.} 
    \label{Fig1_Illustration}  
\end{figure}

The DSS is implemented by mechanically rotating a directional antenna on a turntable to capture signals from various directions, as shown in Fig. \ref{Fig1_Illustration}. Assuming an array for DSS composed of $M$ uniformly distributed elements, the steering angle of the $m$-th element is ${\phi_m} = {2\pi (m - 1)} /M,m \in \left[ {1,M} \right]$. In practice, the rotation step (i.e., ASI) often approximates the HPBW ($\phi_{\rm{3dB}}$) of the directional antenna, ensuring a trade-off between measurement time and DSS accuracy \cite{C2_DSS_FW}, with the number of rotations given by $M = {2\pi }/\phi_{\rm{3dB}}$. It is noteworthy that, due to the relatively slow mechanical rotation speed of the DSS, the measured channel must remain static. The received signal at the $m$-th Rx element is expressed as:
\begin{equation}
\begin{aligned}
    &Y_m(f) = S_m(f) +W_m(f) = H_m(f)U(f) +W_m(f), \\
    &f=[f_1,\dots,f_k,\dots, f_K] \in \left[ f_c-B_w/2,f_c+B_w/2\right],
\end{aligned}
\end{equation}
\noindent
where $f$ represents the frequency components within the range of the center frequency $f_c$ and bandwidth $B_w$. \(S_m(f)\) represents the contribution of the transmitted signal through the channel to the receiver (Rx), and $W_m(f)$ is a circularly symmetric white Gaussian noise component with spectral height $\sigma^2$. $U(f)$ denotes the frequency-domain representation of the transmitted signal, $H_m(f)$ represents the radio channel frequency response (CFR) of the $m$-th element. For simplicity, we confine our discussion to a two-dimensional azimuth plane with an omnidirectional transmitter (Tx) antenna as illustrated in Fig. \ref{Fig1_Illustration}, though the proposed method can be extended to more general three-dimensional scenarios. Therefore, the radio CFR of the $m$-th antenna element is given by:
\begin{equation}
    H_m(f) =g_{{\rm{Tx}}} \sum_{l=1}^{L} \alpha_{l} e^{j\varphi_l}\cdot g_{{\rm{Rx}}}\left(\phi_m-\phi_l\right) \exp\left(-j2\pi f \tau_l\right), 
\end{equation}
\noindent where $L$ is the number of MPCs. The parameters of the $l$-th MPC include the complex amplitude (amplitude $\alpha_l$ and phase $\varphi_l$), the delay $\tau_l$, and the azimuth angle of arrival (AAoA) $\phi_{l}$. $g_{{\rm{Tx}}}$ and $g_{\rm{Rx}} \left(\phi_m-\phi_l\right)$ denote the antenna responses of the Tx omnidirectional antenna and the Rx $m$-th antenna toward the direction $\phi_l$, respectively. The channel impulse response is obtained by applying the inverse Fourier transformation to the CFR:
\begin{equation}
{h_m}\left( {{\tau}} \right) = \frac{1}{{\sqrt K }}\sum\limits_{k=1}^{K} {{H_m}\left( {{f_k}} \right)\exp \left( {j{{2\pi }}{f_k}{\tau}} \right)},
\end{equation}
where $\tau=\left[0,\Delta\tau,2\Delta\tau,\dots,(k-1)\Delta\tau,\dots,(K-1)\Delta\tau\right]$, and $\Delta\tau = 1/B_w$. The corresponding PDP is expressed as:
\begin{equation}
{\rm{PDP}}_m (\tau) =  \left| {h_m}\left( {{\tau}} \right)\right|^2.
\label{equ_delay_power}
\end{equation}

\subsection{Traditional Methods And Drawbacks}
Existing studies typically synthesize the directional PDPs into an omnidirectional PDP and subsequently derive the corresponding MPC angle and power. One method involves selecting direction with the largest peak power from different sounding directions within the same delay bin \cite{C2_Omni_M1_1,C2_Omni_M1_2}, resulting in the synthesized omnidirectional PDP as follows:
\begin{equation}
{\rm{PDP}^{\rm{o-1}}\left( \tau \right)}={\mathop {\max }\limits_m {\rm{PDP}}_m\left( \tau \right)}.
\end{equation} 

The corresponding MPC parameters are expressed as:
\begin{equation}
\begin{aligned}
\mathcal{P}^{\rm{o-1}}  =& \left\{ {\hat{\tau}_l^{\rm{o-1}}}, \ \hat{\phi}_{{m}_l^{\rm{o-1}}},\ \hat{P}^{\rm{o-1}}_l\right\}_{l=1}^{L^{\rm{o-1}}}, \\
\{{\hat{\tau}_l^{\rm{o-1}}}\}&_{l=1}^{L^{\mathrm{o-1}}}  = \mathrm{peak}\left(\mathrm{PDP}^{\mathrm{o-1}}(\tau)\right),\\
 \ {m}_l^{\rm{o-1}} =  \mathop {\arg\max }\limits_m &  {\rm{PDP}}_m({\hat{\tau}_l^{\rm{o-1}}}), \ {\hat{P}}^{\rm{o-1}}_l = {\rm{PDP}}^{\rm{o-1}}({\hat{\tau}_l^{\rm{o-1}}}),
\end{aligned}
\end{equation}
\noindent
where ${L^{\mathrm{o-1}}}$ denotes the number of MPCs estimated by this method, which is fewer than $L$. \(\mathrm{peak}(\cdot)\) is a function for extracting local maxima (peaks) from a given vector. However, the MPC power and AAoA derived from this method exhibit deviations from the true values. On the one hand, because the DSS performs discrete sampling in the spatial domain with a fixed ASI, the MPC's AAoA is quantized to the nearest sampling value, introducing a quantization error, expressed as \(\varepsilon_{\phi} =  \phi_{\rm{3dB}} \cdot \left\lfloor \dfrac{\phi}{\phi_{\rm{3dB}} } + 0.5 \right\rfloor - \phi\). On the other hand, the gain of the directional antenna varies with angle, and it reaches its maximum at the center of the main lobe, while the MPC's AAoA are unlikely to align with this maximum gain direction. When decoupling the antenna response for MPC power calculations, the maximum gain value of the directional antenna is typically subtracted, causing an underestimation of MPC power, represented as \(\varepsilon_P = 10 \log_{10}\left( |g_{\text{Rx}}(0)|^2 - |g_{\text{Rx}}(\phi_l - \phi_m)|^2 \right)\), measured in dB. 

An alternative method involves summing the PDPs from different sounding directions within the same delay bin to produce an omnidirectional PDP with approximately isotropic gain, given by:
\begin{equation}
{\rm{PDP}}^{\rm{o-2}}\left( \tau \right)={\mathop {\sum }\limits_m {\rm{PDP}}_m\left( \tau \right)}.
\end{equation}

Similarly, the parameters of ${L^{\mathrm{o-2}}}$ MPCs are expressed as:
\begin{equation}
\begin{aligned}
\mathcal{P}^{\rm{o-2}} & = \left\{ {\hat{\tau}_l^{\rm{o-2}}}, \ \hat{\phi}_{{m}_l^{\rm{o-2}}},\ \hat{P}^{\rm{o-2}}_l\right\}_{l=1}^{L^{\rm{o-2}}}, \\
\{{\hat{\tau}_l^{\rm{o-2}}}\}&_{l=1}^{L^{\mathrm{o-2}}}  = \mathrm{peak}\left(\mathrm{PDP}^{\mathrm{o-2}}(\tau)\right),\\
 \ {m}_l^{\rm{o-2}} =  \mathop {\arg\max }\limits_m   &{\rm{PDP}}_m({\hat{\tau}_l^{\rm{o-2}}}), \ {\hat{P}}^{\rm{o-2}}_l = {\rm{PDP}}^{\rm{o-2}}({\hat{\tau}_l^{\rm{o-2}}}).
\end{aligned}
\end{equation}

Although this method aims to reduce the error in MPC power (in practice it still depends on how well the HPBW matches the ASI), quantization errors in MPC's AAoA still persist.

\subsection{Proposed Method}
\begin{figure}[htbp]
    \centering  
  \subfloat[]{\includegraphics[width=4.3cm]{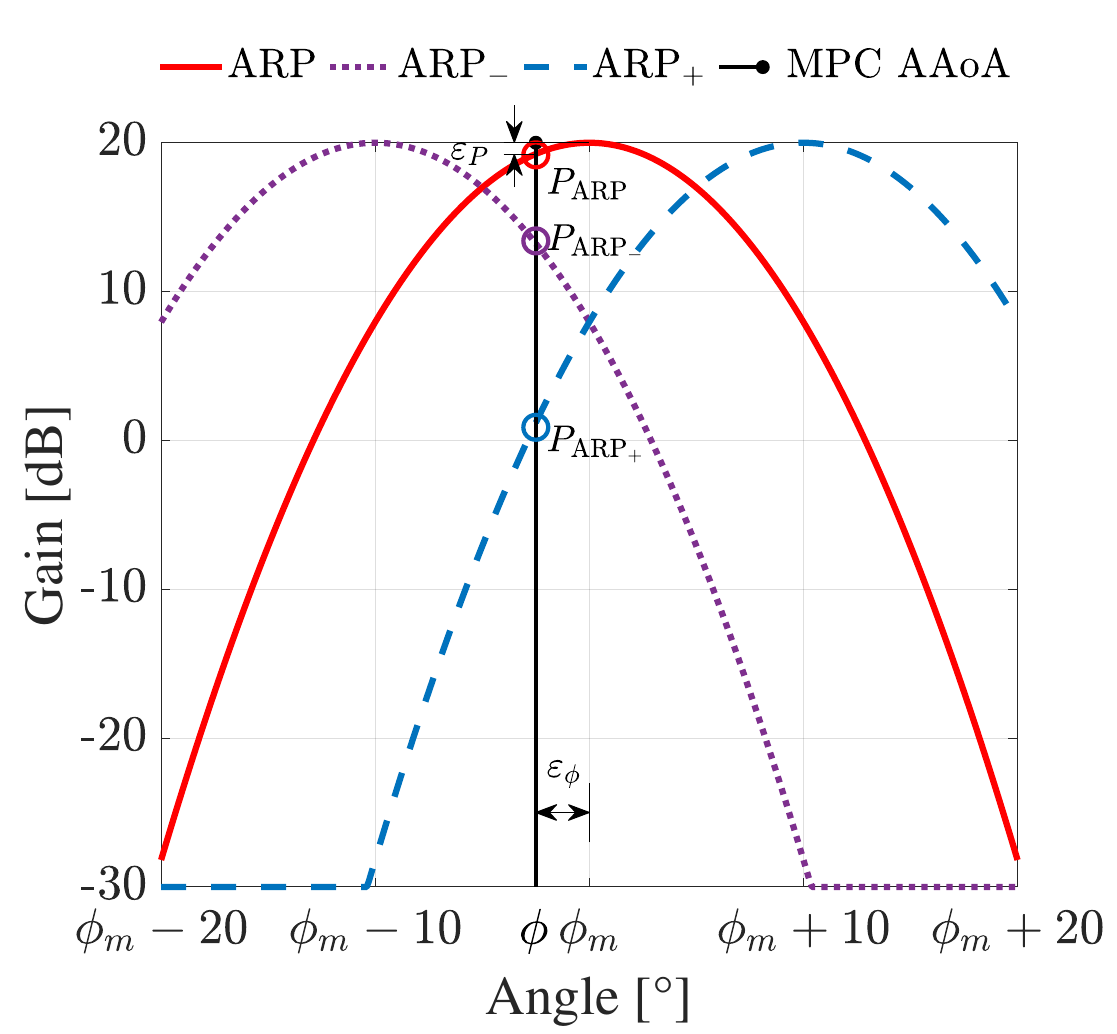} \label{Fig3_Algorithm_diagram}}
  \hfill
  \subfloat[]{\includegraphics[width=4.3cm]{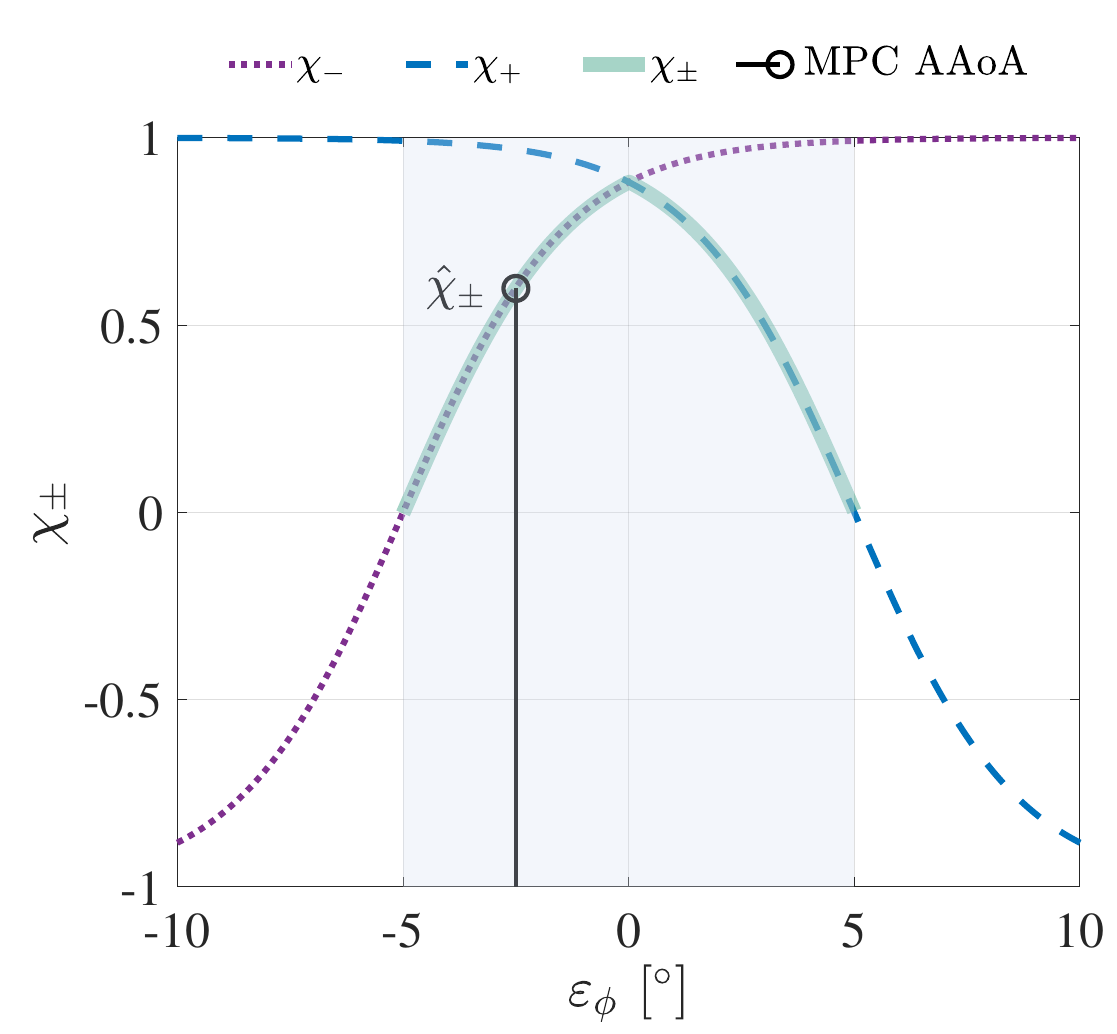} 
  \label{Fig3_Metrics}}
    \caption{(a) ARP in adjacent sounding directions. (b) The relationship between $\chi_\pm$ and the offset angle $\varepsilon _\phi$}
\end{figure}

Traditional methods select only the direction with the highest power within a delay bin for MPC angle estimation and neglect channel responses from adjacent scanning angles. The fundamental concept of our proposed method involves estimating the MPC's AAoA and power by leveraging power differences in adjacent DSS directions. As shown in Fig. \ref{Fig3_Algorithm_diagram}, when the MPC is incident at \(\phi\), the ARP corresponding to the red solid line provides the strongest gain, while adjacent directional antennas (denoted as $\mathrm{ARP}_-$ and $\mathrm{ARP}_+$) also provide slightly smaller gains. Therefore, the directional antenna aligned at \(\phi_m = \phi_{\rm{3dB}} \cdot \left\lfloor \frac{\phi}{\phi_{\rm{3dB}}} + 0.5 \right\rfloor \) receives the maximum power among all elements in the DSS, i.e., \(P_{{\rm{ARP}}} > P_{{\rm{ARP}}_-} > P_{{\rm{ARP}}_+}\). The power difference between them depends on the offset angle \({\varepsilon _\phi}\) of the MPC direction relative to the antenna's maximum reception power direction. Here, a Gaussian beam is employed to generate ARP, which is expressed as:
\begin{equation}
g(\phi) = \sqrt{G_{\max}}  {e^ { \kappa \left(\cos \phi -1\right)}},
\label{equ_von_pattern}
\end{equation}
\noindent where $\kappa=\left({\ln{\sqrt2}}\right)/\left({1-\cos(0.5\phi_{\rm{3dB}})}\right)$ measures the concentration of ARP, which depends on the HPBW of the directional antennas. $G_{\max}$ represents the maximum power gain of the directional antenna. $G_{\max}$ and HPBW are set to 20 dB and 10$^{\circ}$ in this communication. The proposed method consists of two stages: coarse estimation and fine estimation. 

\graphicspath{{picture/}}

\noindent
\textbf{Stage 1}: Coarse estimation:

First, the PDPs measured from each direction are combined into a PADP, $ \mathcal{A} \in \mathbb{C}^{M \times K}$, expressed as:
\begin{equation}
\begin{aligned}
     \mathcal{A}\left( m,k \right) & = \left[ {\rm{PDP}}_1 \left( \tau \right)^T, \cdots ,{\rm{PDP}}_m \left( \tau \right)^T, \cdots ,{\rm{PDP}}_M \left( \tau \right)^T \right]^T ,
\end{aligned}
\end{equation}

Then, local maxima are searched within the two-dimensional PADP space to obtain the set of coarse MPC parameters ${\cal P}^{(1)}$ as given in (\ref{equ_coarse_estimation}). The condition \(\mathcal{A}(m,k) > \mathcal{A}(m',k') \ \forall (m',k') \in \mathcal{N}(m,k)\) ensures that the position \((m,k)\) is a local maximum in its neighborhood \(\mathcal{N}(m,k)\). Since the angle is periodic, the adjacent angle indices for the \(m\)-th position are denoted as \(m_-\) and \(m_+\). Unlike traditional methods, which identify peaks in the one-dimensional delay domain after synthesizing the omnidirectional PDP, the HAED method finds peaks in the two-dimensional PADP space. This method effectively avoids the loss of MPCs with different AAoAs but the same delay bin. For example, in symmetric scenarios such as corridors, two MPCs from opposite side walls typically have the same delay but different AAoAs. Simple omnidirectional synthesis often results in the loss of one MPC.
\begin{figure*}[b]
\hrulefill
\begin{align}
&{\cal P}^{(1)} = \left\{ {\left. {\left( {{{\hat \tau }_l} = (k - 1)\Delta \tau ,\hat \phi _l^{(1)} = {\phi _m},\hat P_l^{(1)} = {\cal A}(m,k)} \right)} \right|{\cal A}(m,k) > {\cal A}(m',k')\;\forall (m',k') \in {\cal N}(m,k)} \right\}_{l = 1}^L, \label{equ_coarse_estimation} \\
\mathcal{N}(m,k) = &\left\{ {\left. {\left( {m{\rm{',}}k{\rm{'}}} \right)} \right|m' \in \{ {m_ - },{m_ + }\} ,k' \in \{ k - 1,k + 1\},1 \le k' \le K} \right\},m_- = \left( \left( {m - 2} \right)  {\rm{mod}} \ M \right) + 1, \, m_+ = \left( {m } \ {\rm{mod}} \  M \right)+ 1, \notag
\end{align}
\end{figure*}

\noindent
\textbf{Stage 2}: Fine estimation:

The coarse estimation result defines the range of AAoA as $\left[ {\hat \phi _l^{(1)} - \dfrac{{{\phi_{\rm{3dB}}}}}{2}, \hat \phi _l^{(1)} + \dfrac{{{\phi_{\rm{3dB}}}}}{2}} \right]$. Subsequently, it is necessary to determine the offset angle \(\varepsilon_\phi\) of the AAoA relative to the sampling angle \(\phi_l^{(1)}\). 
We have observed a dependency between the power difference received by adjacent antennas in DSS and the MPC offset angle. Therefore, a normalized power difference metric \cite{Dai_Metrics}, \(\chi_{\pm}(\varepsilon _\phi)\), is introduced as: 
\begin{equation} 
\chi_{\pm} \left( \varepsilon_\phi \right) = \dfrac{ g^2 \left(\varepsilon_\phi \right)  -  g^2 \left(\varepsilon_\phi \mp \phi_{\rm{3dB}}\right) }{ g^2 \left(\varepsilon_\phi \right)  +  g^2\left(\varepsilon_\phi \mp \phi_{\rm{3dB}}\right) },
\end{equation}
where \(\chi_{-}\) and \(\chi_{+}\) represent the power differences between the ARP and $\text{ARP}_-$ pair, and between the ARP and $\text{ARP}_+$ pair, respectively. When the received power of $\text{ARP}_-$ is greater than that of $\text{ARP}_+$, the \( \pm \) in \( \chi_{\pm} \) is assigned as \( - \), with the corresponding \( \mp \) assigned as \( + \), indicating the first ARP pair. Otherwise, \( \pm \) is assigned to \( + \), and \( \mp \) is assigned as \( - \), representing the second pair. By applying (\ref{equ_von_pattern}), \(\chi_{\pm}(\varepsilon _\phi)\) can be rewritten as: 
\begin{equation} 
\chi_{\pm} \left( \varepsilon_\phi \right) = \frac{ e^{2\kappa (\cos \varepsilon_\phi - 1)} - e^{2\kappa (\cos (\varepsilon_\phi \mp \phi_{\rm{3dB}}) - 1)}}{ e^{2\kappa (\cos \varepsilon_\phi - 1)} + e^{2\kappa (\cos (\varepsilon_\phi \mp \phi_{\rm{3dB}}) - 1)}}.
\end{equation}

By inverting the function \(\chi\left( \cdot \right)\), the offset angle \({\varepsilon _\phi}\) can be explicitly estimated as: 
\begin{equation}
    \varepsilon_\phi = \arcsin \left( \frac{\mp 1}{4\kappa\, \sin \left( \dfrac{ \phi_{\rm{3dB}}}{2} \right)} \ln \left( \frac{ \chi_{\pm} + 1}{1 -  \chi_{\pm}} \right) \right) \pm \frac{\phi_{\rm{3dB}}}{2}.
    \label{equ_phi}
\end{equation}

Fig. \ref{Fig3_Metrics} illustrates the relationship between the offset angle \({\varepsilon _\phi}\) and the normalized power differences \(\chi_{\pm}\), demonstrating that \(\hat\chi_{\pm}\) and \(\varepsilon_\phi\) are uniquely correlated. Consequently, the unique offset angle \(\varepsilon_\phi\) of the MPC can be accurately estimated. For the measured data, \(\hat \chi_{\pm,l}\) can be calculated based on the coarse estimated power \(\hat{P}_l^{(1)}\), expressed as: 
\begin{equation} 
\begin{aligned}
\begin{array}{c}
{\hat \chi _{\pm,l}} = \dfrac{{{\hat{P}}_l^{(1)} - {P_{l\mp}}}}{{{\hat{P}}_l^{(1)} + {P_{l\mp}}}}, \; \; {m_l} = \dfrac{{M\hat \phi _l^{(1)}}}{{2\pi }} + 1,\\
{P_{l-}} = {\rm{PD}}{{\rm{P}}_{({m_{l})_-}}}({{\hat{\tau}} _l}),\;\;\;{\kern 1pt} {P_{l+}} = {\rm{PD}}{{\rm{P}}_{({m_{l}})_+}}({{\hat{\tau}} _l}),
\end{array}
\end{aligned}
\end{equation}
\noindent
where \(m_l\) denotes the index of the directional antenna with the maximum received power for the \(l\)-th MPC, while \((m_{l})_-\) and \((m_{l})_+\) represent the indices of the adjacent elements to \(m_l\). Similarly, \({P_{l_-}}\) and \({P_{l_+}}\) correspond to the received power of the \(l\)-th MPC at the adjacent elements. When the received power of \(\mathrm{ARP}_-\) is greater than that of \(\mathrm{ARP}_+\), the power \(P_{l_-}\) and the normalized metric \(\chi_-\) are selected; otherwise, \(P_{l_+}\) and \(\chi_+\) are chosen. By substituting \(\hat{\chi}_{\pm,l}\) into (\ref{equ_phi}), the corresponding offset angle \(\varepsilon_{\phi,l}\) can be calculated, enabling correction of the MPC power. For practical directional antennas, a closed-form expression like (\ref{equ_von_pattern}) may not exist. However, the discretized \({\chi _ \pm }({{\varepsilon _\phi }})\) can be obtained based on the ARP calibrated in an anechoic chamber, enabling the estimation of \({\varepsilon _{\phi ,l}}\) by minimizing the absolute error between \({\hat \chi _{ \pm ,l}}\) and \({\chi _ \pm }\), expressed as:
\begin{equation}
\begin{aligned}
{\varepsilon _{\phi ,l}} = \mathop {\arg \min }\limits_{{\varepsilon _\phi }} \left( {\left| {{{\hat \chi }_{ \pm ,l}} - {\chi _ \pm }\left( {{\varepsilon _\phi }} \right)} \right|} \right).
\end{aligned}
\end{equation}

The parameters of the MPC, estimated using the HAED method, can then be expressed as:
\begin{equation}
\begin{aligned}
{\cal P}^{\rm{HAED}} = &  \left\{ { {\hat{\tau}}_l, \hat{\phi}_l,\hat{P}_l } \right\}_{l=1}^L, \\
\hat{\phi_l} = \left(\hat{\phi}^{(1)}_l + \varepsilon_{\phi,l}\right) \, {\rm{mod}} &\ 2\pi, \;
\hat{P_l} = \hat{P}^{(1)}_l  \frac{{g^2\left(0\right)}}{{g^2\left(\varepsilon_{\phi,l} \right)}}.
\end{aligned}
\end{equation}

Compared to traditional methods, the proposed method significantly enhances the accuracy of MPC angle estimation without increasing the DSS time.

\section{Simulation Validation}
\label{sec_3}

This section evaluates the performance of traditional MPC angle estimation methods and the proposed HAED method from two aspects: numerical simulation (i.e., CRLBs) and ray-tracing (RT) simulation.

\subsection{Derivation of CRLBs for performance evaluation}

Since the square root of the CRLB provides a lower bound on the root mean square estimation error (RMSEE), the estimation performance of $\hat \phi$ and $\hat \alpha$ can be evaluated using $\sqrt{{\text{CRLB}}({\hat \phi})}$ and $\sqrt{{\text{CRLB}}({\hat \alpha}/\alpha)}$, respectively \cite{Kim_CRLB}. Based on the signal model described in Section \ref{sec_2}, the log-likelihood function of the wideband DSS can be expressed as:
\begin{equation}
   \Lambda(\boldsymbol{\Theta}; \boldsymbol{Y}) = - MK\ln \left( {\pi {\sigma ^2}} \right) - \frac{1}{{{\sigma ^2}}}{\left\| {{\boldsymbol{Y}} - {\boldsymbol{S}}\left( {\bf{\Theta }} \right)} \right\|^2} , 
\end{equation}
where ${\boldsymbol{\Theta }}=\left[{\hat{\boldsymbol{\theta }}}_1,\dots,{\hat{\boldsymbol{\theta }}}_L \right]$ denotes the set of parameters to be estimated for $L$ MPCs, with ${\hat{\boldsymbol{\theta }}}_l = \left[ \alpha_l,\varphi_l,\phi_l,\tau_l \right]$. $\boldsymbol{Y}\in{\mathbb{C}}^{MK\times 1}$ and $\boldsymbol{S}(\boldsymbol{\Theta}) \in \mathbb{C}^{MK \times 1}$ are defined as:
\begin{equation}
\begin{aligned}
       \boldsymbol{Y} &= {\rm{vec}} \left[ Y_m \left( f\right);\ f=f_1,\dots,f_K, m = 1,\dots,M\right] , \\
       {\boldsymbol{S}}\left( {\bf{\Theta }} \right) &= {\rm{vec}} \left[S_m \left( f\right) ;\ f=f_1,\dots,f_K, m = 1,\dots,M\right] , 
\end{aligned}
\end{equation}
where ${\text{vec}}[\cdot]$ denotes the vectorization operation. For an unbiased estimator \(\hat{\boldsymbol{\Theta }}\), its mean squared estimation error lower bound, known as the CRLB, is given by \cite{CRLB_book}:
\begin{equation}
  E\left[ {{{\left( {{\bf{\hat \Theta }} - {\bf{\Theta }}} \right)}^H}\left( {{\bf{\hat \Theta }} - {\bf{\Theta }}} \right)} \right] \ge {{\boldsymbol{F}}^{ - 1}}\left( {\bf{\Theta }} \right) = {\rm{CRLB}}\left( {{\bf{\hat \Theta }}} \right),
\end{equation}
\noindent
where ${\boldsymbol{F}}\left( {\boldsymbol{\Theta }} \right)$ denotes Fisher’s information matrix (FIM). The element in the \(i\)-th row and \(j\)-th column of \({{\boldsymbol{F}}}\left( {\boldsymbol{\Theta }} \right)\) can be expressed as \cite{SAGE}: 
\begin{equation}
   {\boldsymbol{F}}_{ij}(\boldsymbol{\Theta}) = -{E} \left[ \frac{\partial}{\partial \boldsymbol{\Theta}_i} \frac{\partial}{\partial \boldsymbol{\Theta}_{j}} \Lambda(\boldsymbol{\Theta}; \mathbf{y}) \right], 
   \label{equ_FIM}
\end{equation}
where $E[\cdot]$ represents the expectation operator. By substituting the ARP from (\ref{equ_von_pattern}), the CRLBs for each parameter of the \(L\) MPCs can be calculated, which are the diagonal elements of the inverse FIM. However, since the Fisher matrix may be non-diagonal, deriving an explicit CRLB expression for the multipath case is challenging. In this communication, expressions for the CRLBs is derived for the special case of \(L = 1\), and simulation results for the CRLBs with \(L = 2\) are presented. In the subsequent simulations, the center frequency is set to \( f_c = 37.5 \, \text{GHz} \), bandwidth \( B_w = 2 \, \text{GHz} \), number of frequency points \( K = 1001 \), and \( U(f) = \sqrt{P_u} \). The directional ARP is configured with a HPBW of \( \phi_{\rm{3dB}} = 10^\circ \) and a maximum power gain of 20 dB, i.e., \( G_{\rm{max}} = 20 \ {\rm{dB}}\). Additionally, the RMSEE of the MPC's AAoA and amplitude is assessed through Monte Carlo simulations. A total of 1000 Monte Carlo simulations are performed for each signal-to-noise ratio (SNR).

\subsubsection{Special Case-Estimation of One MPC}
\begin{figure}
  \centering
  \subfloat[]{\includegraphics[width=4.3cm]{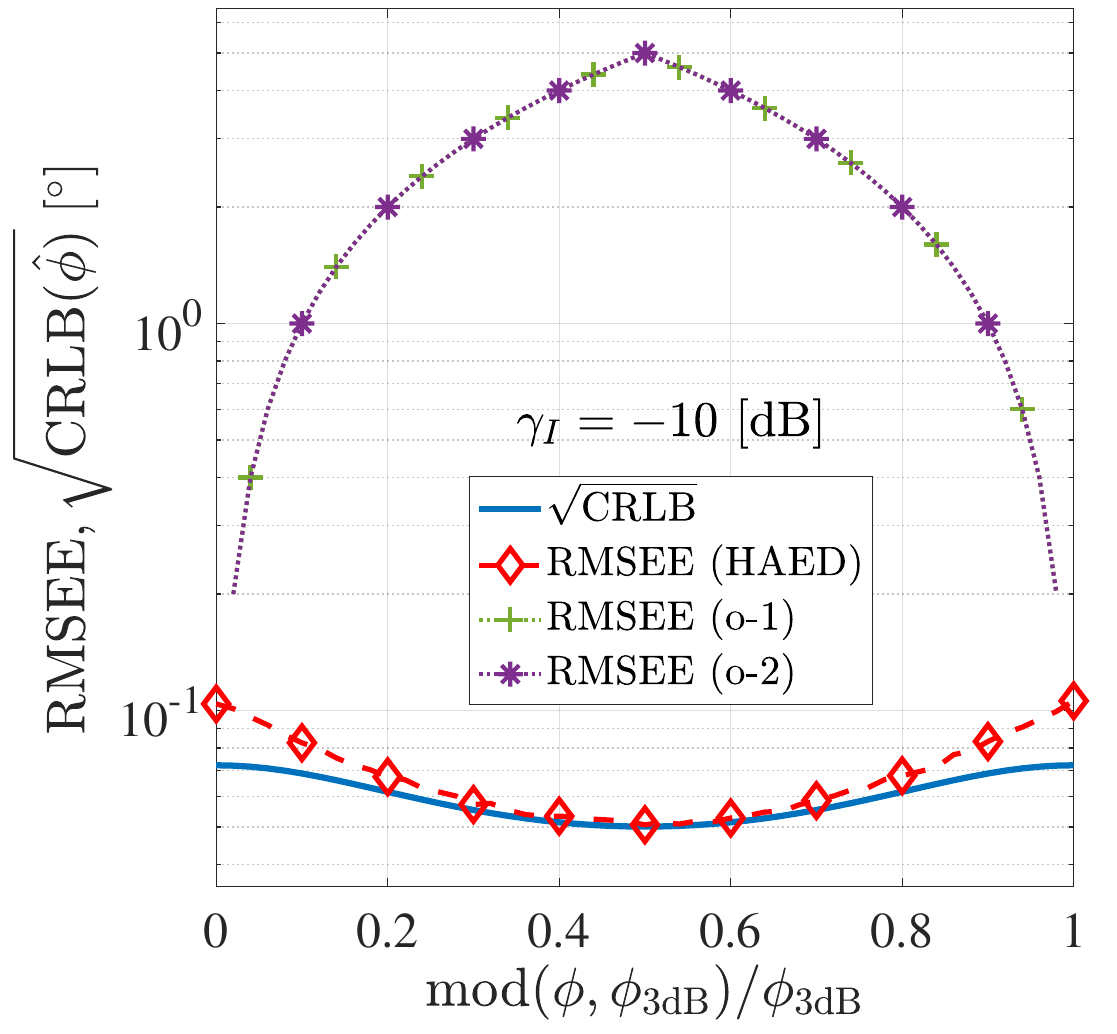} \label{Fig_CRLB1_Angle_phi}}
  \hfill
  \subfloat[]{\includegraphics[width=4.3cm]{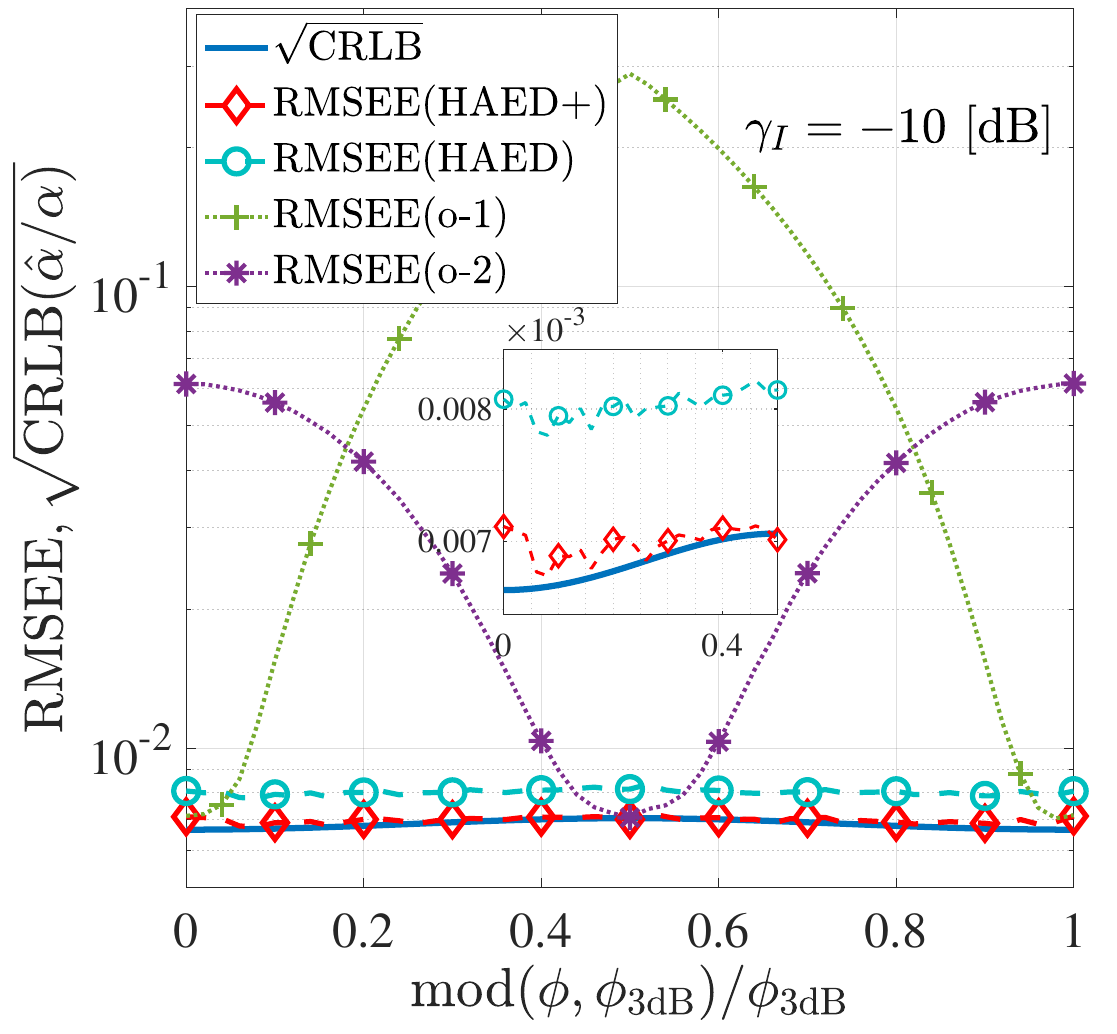} 
  \label{Fig_CRLB1_Angle_alpha}}
  \\[-3.5mm]
  \subfloat[]{\includegraphics[width=4.3cm]{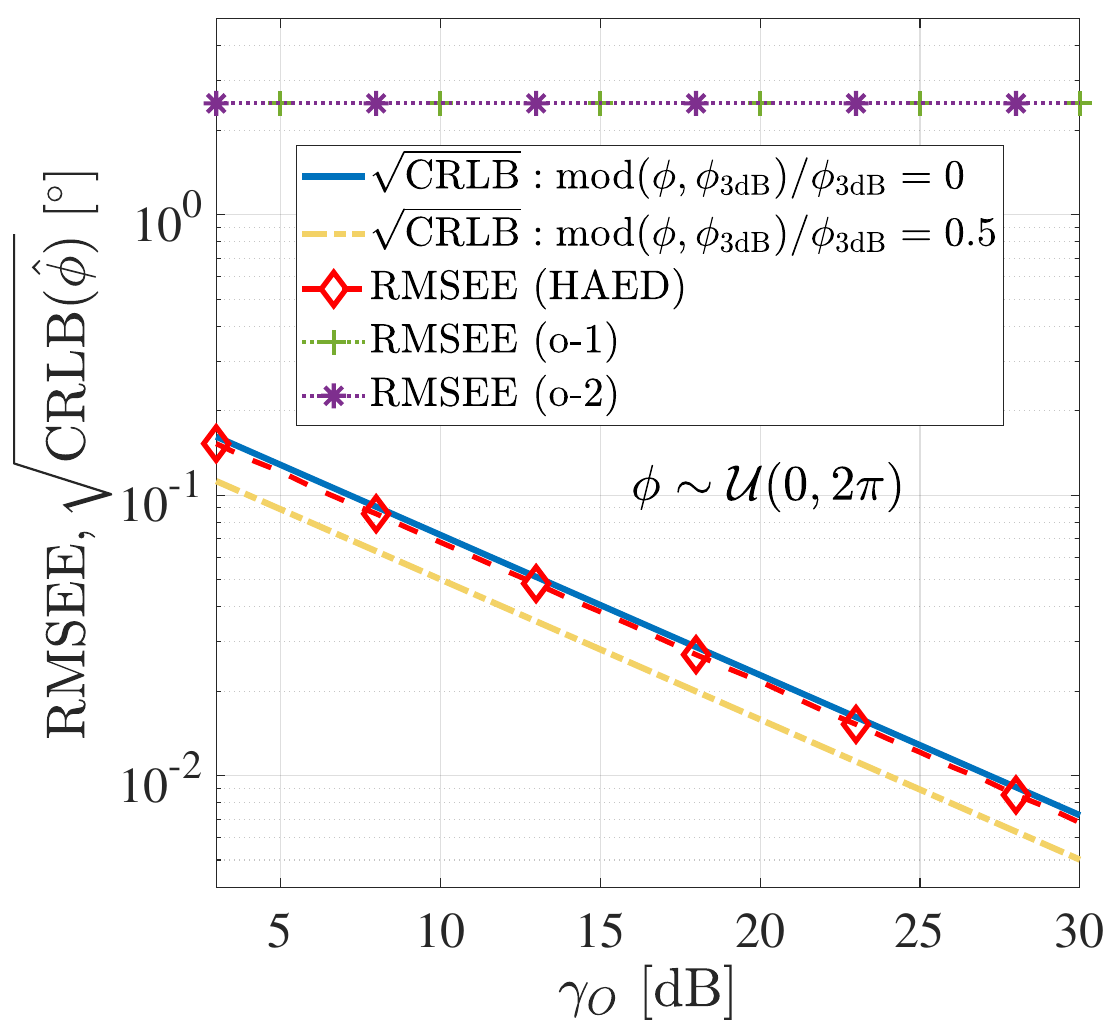} \label{Fig_CRLB1_SNR_phi}}
  \hfill
  \subfloat[]{\includegraphics[width=4.3cm]{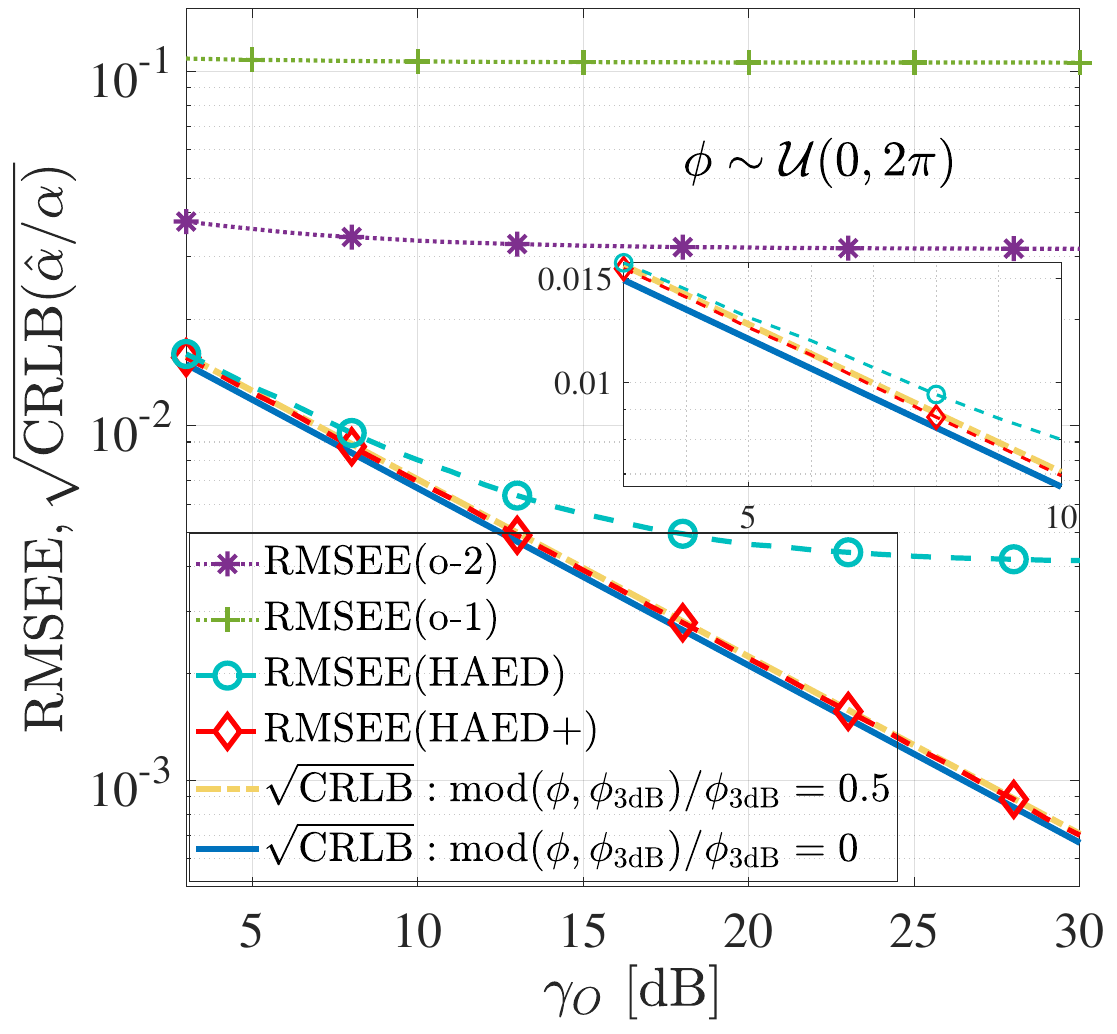} \label{Fig_CRLB1_SNR_alpha}}
  \caption{The variation of $\sqrt{\rm{CRLB}}$ and RMSEE with MPC angle for (a) AAoA and (b) normalized amplitude, and the variation of $\sqrt{\rm{CRLB}}$ and RMSEE with output SNR for (c) AAoA and (d) normalized amplitude.}
  \label{Fig_CRLB1}
\end{figure}

When there is only a single MPC, the parameters are uncoupled, so the FIM \( {\boldsymbol{F}}\left( {\boldsymbol{\Theta }} \right) \) is a diagonal matrix. As a result, the expressions for the CRLBs can be easily derived as:
\begin{equation}
{\rm{CRLB}}\left( {\hat \phi } \right) = \frac{1}{{2{\gamma _I}K{\kappa ^2}\sum\limits_{m = 1}^M {{{\sin }^2}\left( {{\phi _m} - {\phi _l}} \right){g^2}\left( {{\phi _m} - {\phi _l}} \right)} }},
\end{equation}
\begin{equation}
{\rm{CRLB}}\left( {\frac{{\hat \alpha }}{\alpha }} \right) = \frac{1}{{2{\gamma _I}K\sum\limits_{m = 1}^M {{g^2}\left( {{\phi _m} - {\phi _l}} \right)} }} = {\rm{CRLB}}\left( {\hat \varphi } \right),
\end{equation}
where ${\gamma _I}={ {{{\alpha ^2}} P_u /{{\sigma ^2}}}}$ denotes theinp ut SNR. Figs. \ref{Fig_CRLB1_Angle_phi} and \ref{Fig_CRLB1_Angle_alpha} illustrate the relationships between the \( \sqrt{\rm{CRLB}({\hat{\phi}})}\) and \( \sqrt{\rm{CRLB}({\hat{\alpha}}/\alpha)}\) with the AAoA, respectively. It is evident that the theoretical \( \sqrt{\rm{CRLB}} \) closely matches the empirical RMSEE, both of which are significantly lower than the estimation error of traditional methods. However, signal broadening due to finite bandwidth and discretization errors in sampling result in the RMSEE\( ({\hat{\alpha}}/\alpha) \) being slightly higher than \( \sqrt{{\rm{CRLB}} ({\hat{\alpha}}/\alpha)}\), but still much smaller than the estimation errors of traditional methods. Further improvement in the estimation accuracy of RMSEE\( ({\hat{\alpha}}/\alpha) \) can be achieved through sinc function interpolation in the delay domain, referred to as the HAED+ method, bringing it closer to \( \sqrt{\rm{CRLB}({\hat{\alpha}}/\alpha)} \). The detailed processing steps are omitted here.

Figs. \ref{Fig_CRLB1_SNR_phi} and \ref{Fig_CRLB1_SNR_alpha} simulate the performance of the proposed method under different SNRs. For each SNR, the average RMSEE for different AAoA, which follow a uniform distribution, is provided. Additionally, the \( \sqrt{\rm{CRLB}} \) is presented for \( \mod( \phi,\phi_{\rm{3dB}}) / \phi_{\rm{3dB}} = 0 \) and \(\mod( \phi,\phi_{\rm{3dB}}) / \phi_{\rm{3dB}} = 0.5 \), which represent the minimum and maximum values of \( \sqrt{\rm{CRLB}} \) for different AAoA, as shown in Figs. \ref{Fig_CRLB1_Angle_phi} and \ref{Fig_CRLB1_Angle_alpha}. It is worth noting that the horizontal axis represents the output SNR \( \gamma_O \), which is the input SNR multiplied by the antenna gain, i.e., \( \gamma_O = G_{\rm{max}} \gamma_I \). It can be observed that as the output SNR increases, both \( \sqrt{\rm{CRLB}} \) and the average RMSEE decrease, with the average RMSEE closely approximating \( \sqrt{\rm{CRLB}} \). The estimation accuracy is significantly higher than that of traditional methods. 

These findings indicate that, the further the AAoA deviates from the ARP’s maximum-gain direction, the more the AAoA estimation accuracy improves, whereas amplitude estimation accuracy decreases. As the SNR increases, both AAoA and amplitude estimation accuracy improve. Compared to the two traditional omnidirectional synthesis-based MPC angle estimation methods, the proposed method improves estimation accuracy by an order of magnitude.

\subsubsection{Special Case-Estimation of Two MPCs with Equal Power}

\begin{figure}
  \centering
  \subfloat[]{\includegraphics[width=4.3cm]{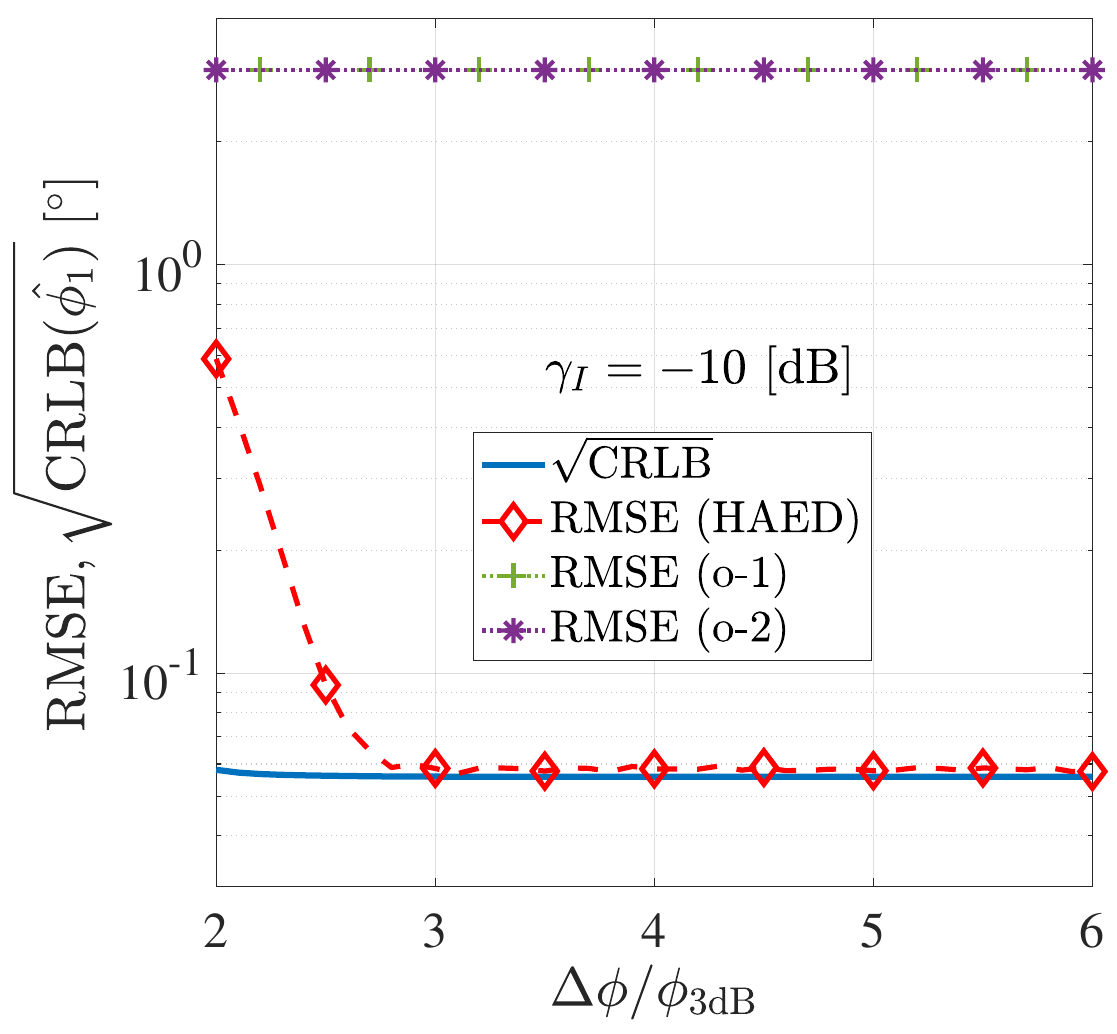} \label{Fig_CRLB2_Angle_phi1}}
  \hfill
  \subfloat[]{\includegraphics[width=4.3cm]{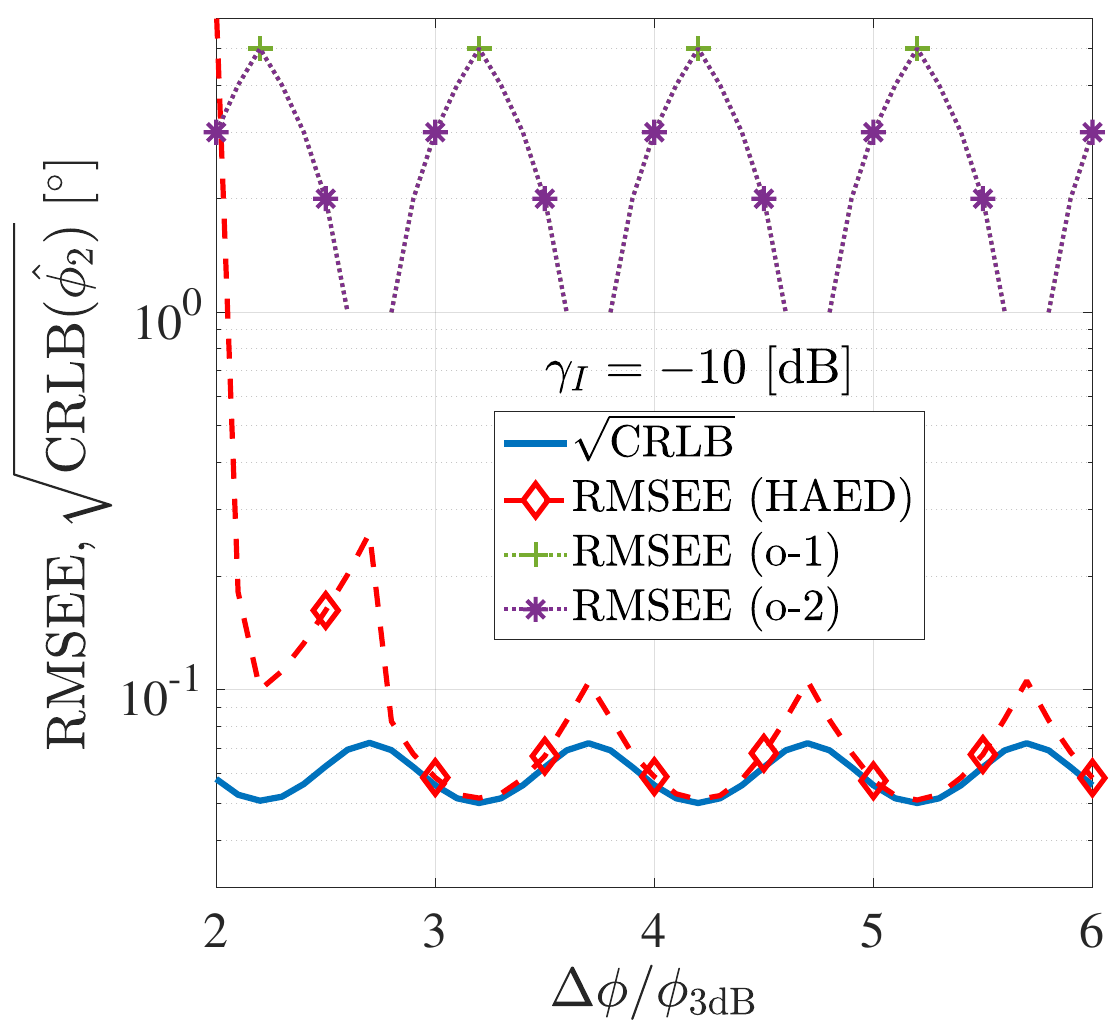} 
  \label{Fig_CRLB2_Angle_phi2}}
  \\[-3.5mm]
  \subfloat[]{\includegraphics[width=4.3cm]{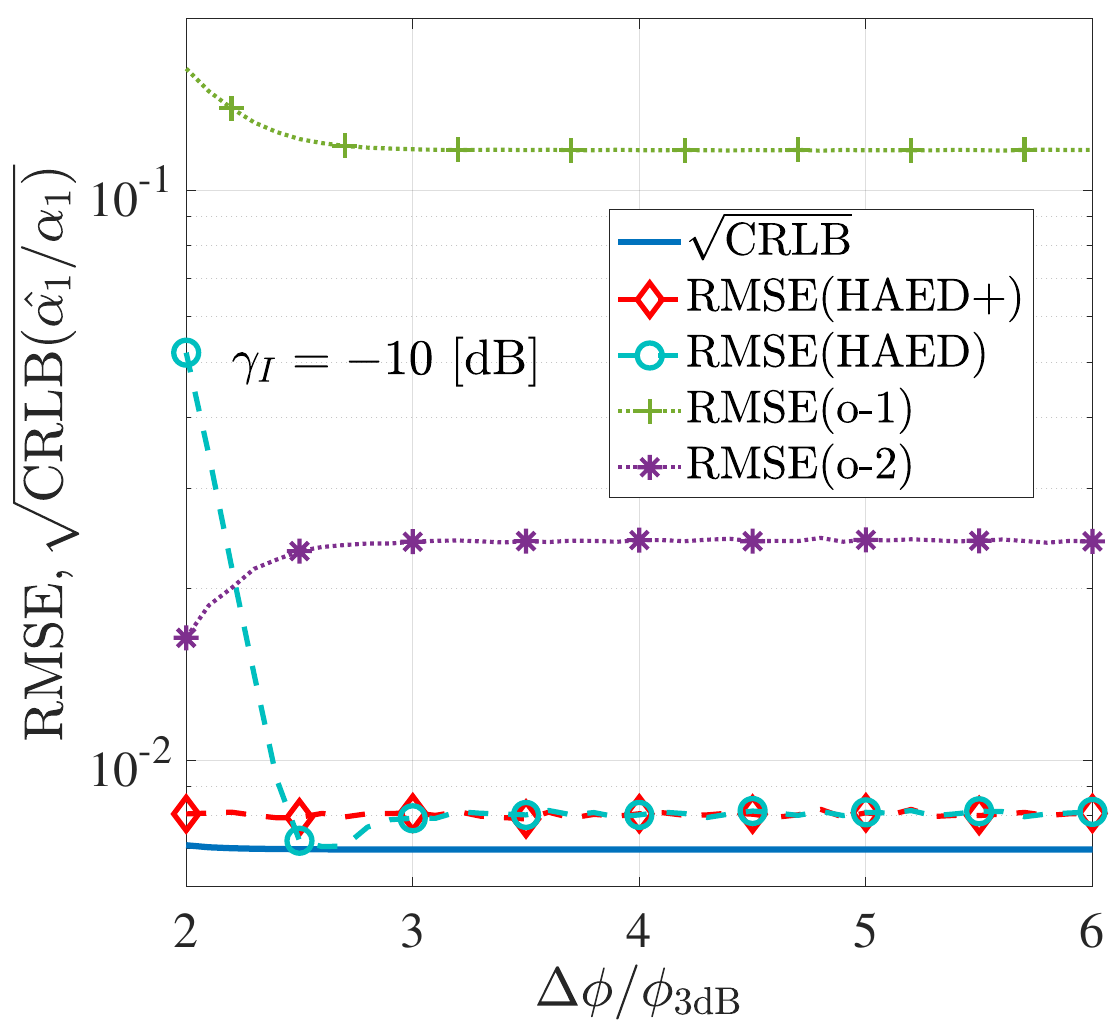} \label{Fig_CRLB2_Angle_alpha1}}
  \hfill
  \subfloat[]{\includegraphics[width=4.3cm]{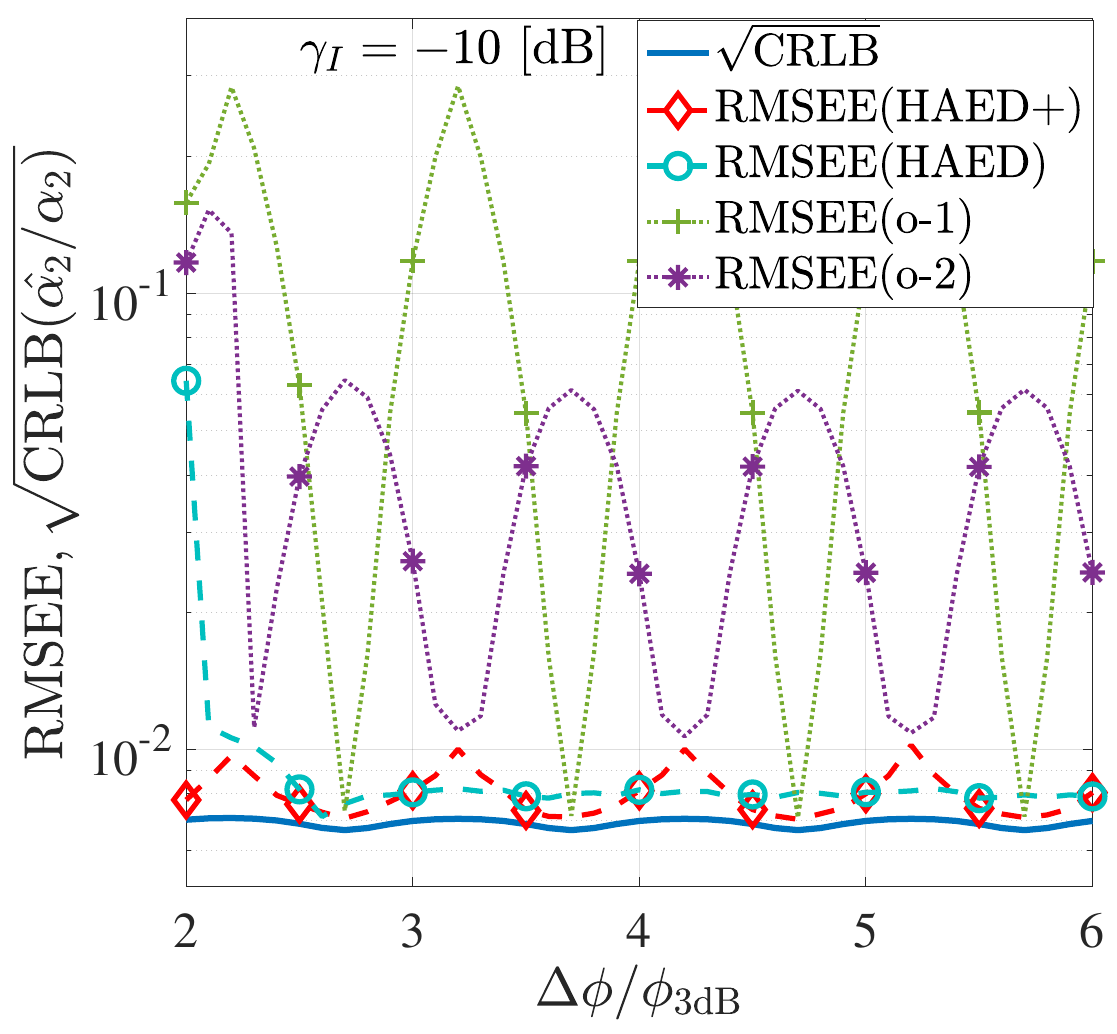} \label{Fig_CRLB2_Angle_alpha2}}
  \caption{The variation of $\sqrt{\rm{CRLB}}$ and RMSEE for two MPC's AAoA with the MPC angular separation, (a) $\phi_1$, and (b) $\phi_2$. The variation of $\sqrt{\rm{CRLB}}$ and RMSEE for two MPC normalized amplitudes with the MPC angle separation, (c) $\hat \alpha_1/\alpha_1$, and (d) $\hat \alpha_2/\alpha_2$.}
  \label{Fig_CRLB2}
\end{figure}

Similarly, the \( \sqrt{\rm{CRLB}} \) for the case of two MPCs can be derived from the FIM as given in (\ref{equ_FIM}). Fig. \ref{Fig_CRLB2} presents the \( \sqrt{\rm{CRLB}} \) and RMSEE for two equal-power, identical-delay MPCs, where \( \phi_1 = 3^{\circ} \), \( \phi_2 = \Delta \phi + \phi_1 \), \( \tau_1 = \tau_2 = 25 \, \text{ns} \), \( \gamma_I = -10 \, \text{dB} \), \( \varphi_1 = \pi/3 \), \( \varphi_2 = \pi/5 \), and \( \alpha_1 = \alpha_2 = 1 \). It is evident that the proposed method results in estimation errors much lower than those of traditional methods, thereby indicating better performance. Moreover, it is observed that when the angular separation between the two MPCs exceeds three times the HPBW, the RMSEE approaches \( \sqrt{\rm{CRLB}} \), implying that the two MPCs can be fully separated. The reason is that when the angular difference between the two MPCs is small, their mutual interference makes it difficult to distinguish them.

In summary, the proposed method requires that MPCs be resolvable in the delay-angle domain. This condition is easily achievable due to the large bandwidth and narrow beams in mmWave and THz channel measurements \cite{C3_LXM}. Specifically, the proposed HAED method satisfies at least one of the following two constraints:
\begin{itemize}
    \item Each delay bin contains only one MPC, i.e., $|\tau_l - \tau_{l'}| > 1/{B_w}, l \neq l'$.
    \item The AAoA difference between two MPCs exceeds three times the HPBW, i.e., $\min(|\phi_l - \phi_{l'}|,2\pi - |\phi_l - \phi_{l'}|) > 3 \phi_{\rm{3dB}}, l \neq l'$.
\end{itemize}

\subsection{Ray-tracing simulation}

RT is a key approach for channel simulation. Therefore, we conducted DSS-based RT simulations to validate the proposed method's accuracy. The simulation was performed in a room with dimensions \(8.34 \times 6.05 \times 2.40 \ \text{m}^3\) at a center frequency of 37.5 GHz. The directional ARP and ASI settings remained consistent with those of the previous numerical simulations. A total of 700 Tx-Rx pairs were deployed, resulting in 133,328 MPCs. 

\begin{figure}
  \centering
  \subfloat[]{\includegraphics[width=4.3cm]{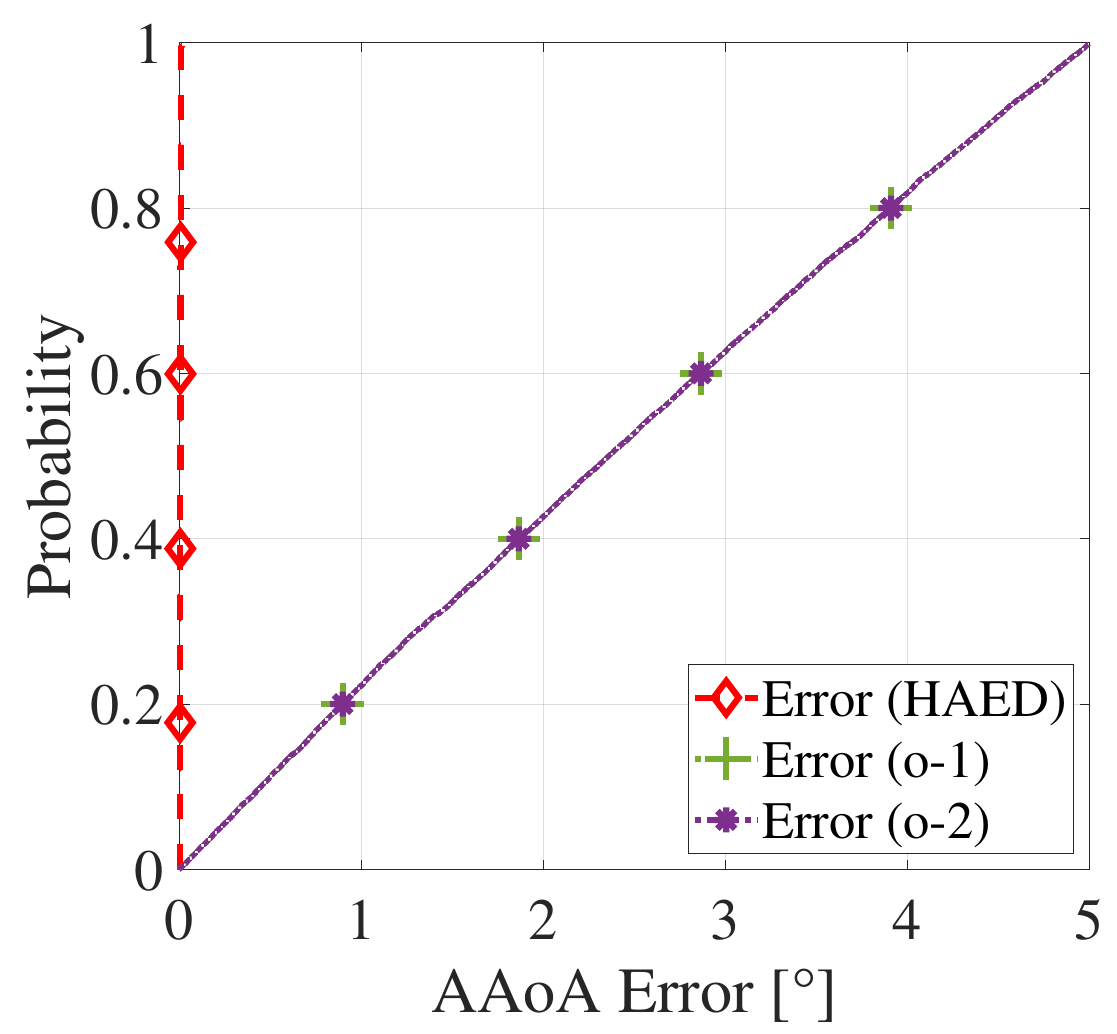} \label{Fig_RT_AAoA}}
  \hfill
  \subfloat[]{\includegraphics[width=4.3cm]{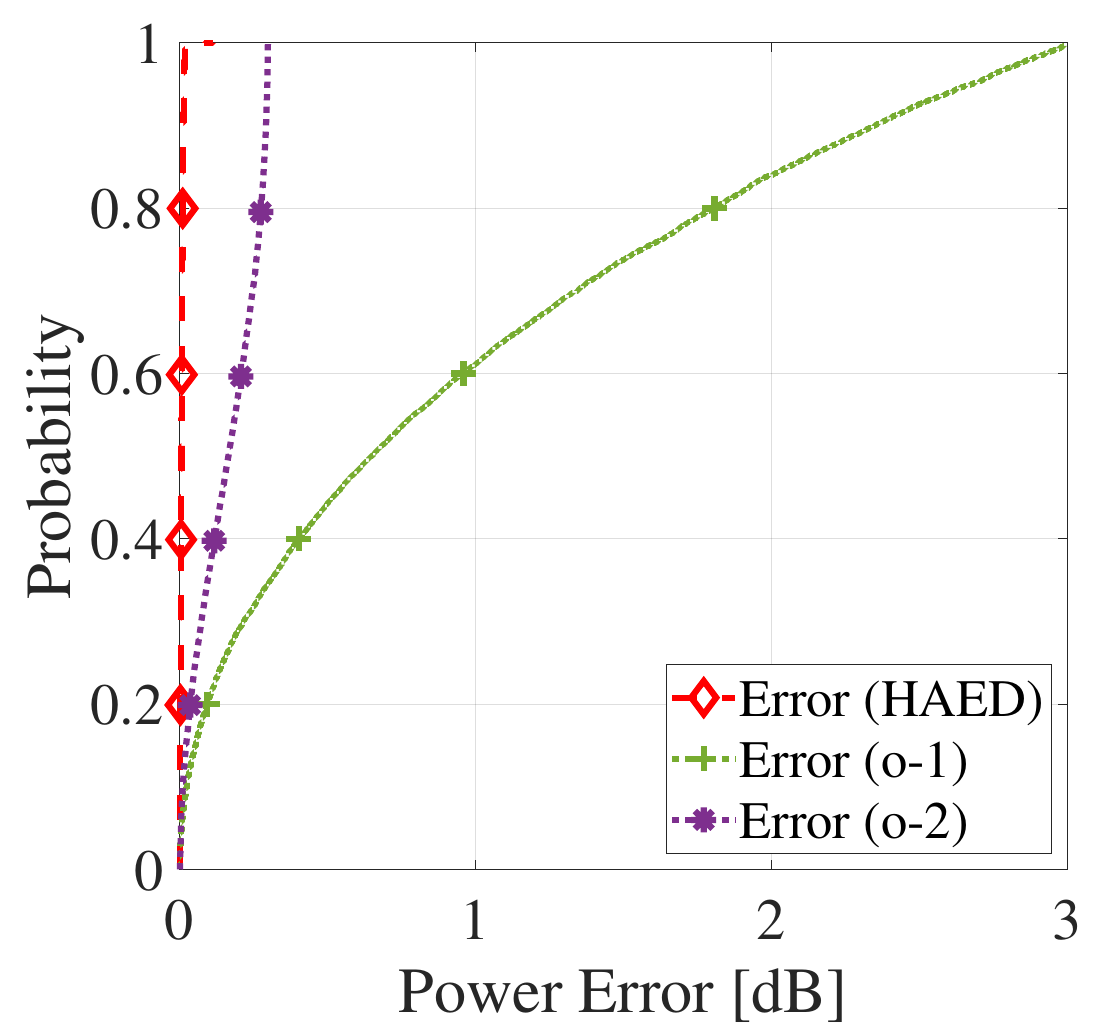} \label{Fig_RT_Power}}
  \caption{The estimation errors of MPC parameters for three methods: (a) AAoA, (b) Power.}
  \label{Fig_RT}
\end{figure}

Fig. \ref{Fig_RT} presents the cumulative distribution function of the estimation errors for the AAoA and power of these MPCs. For AAoA estimation, the traditional method's accuracy is constrained by the DSS's ASI, with errors approximately uniformly distributed between 0° and 5°, and an average error of 2.39°. In contrast, the proposed method achieves more accurate AAoA estimation. Regarding MPC power estimation, the simulation was set with ASI = HPBW. In the traditional o-1 method, the estimation error ranges from 0 to 3 dB, with an average power error of 0.93 dB. The traditional o-2 method compensates for part of the error by utilizing the received power in adjacent detection directions, but due to the imperfect omnidirectional gain synthesis of DSS, an average error of 0.16 dB remains. Conversely, the proposed method estimates the MPC power with near-perfect accuracy. In summary, the RT simulation results demonstrate that, compared to traditional methods, the proposed method significantly improves the estimation accuracy for both MPC's AAoA and power.

\section{Measurement Validation}
\label{sec_4}
\subsection{Measurement Campaign}
\begin{figure}
  \centering
  \subfloat[]{\includegraphics[width=4cm,height=5cm]{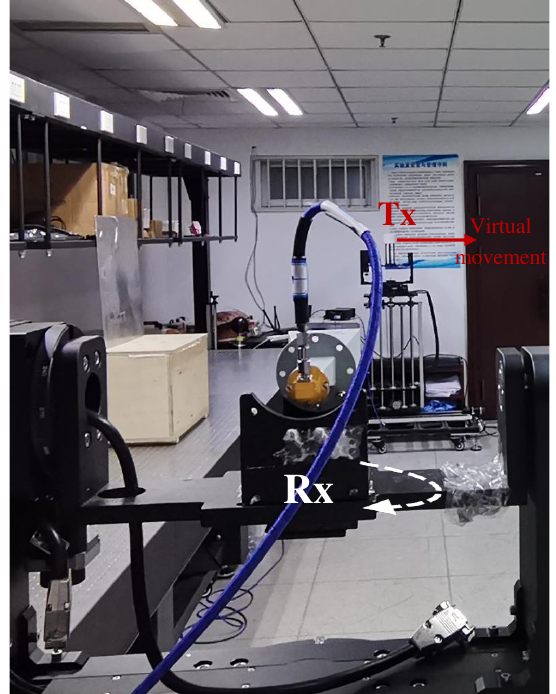} \label{Fig_Mea_scenario_a}}
  \hfill
  \subfloat[]{\includegraphics[width=4cm,height=5cm]{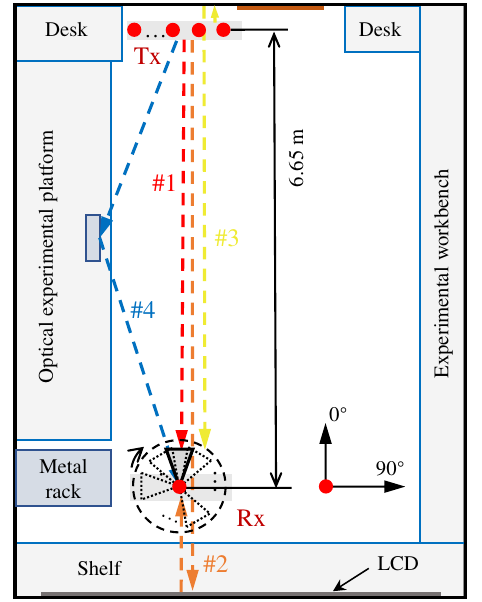} \label{Fig_Mea_scenario_b}}
  \caption{(a) The photograph and (b) the geometric layout of the measurement scenario.}
  \label{Fig_Mea_scenario}
\end{figure}

The measurement scenario is identical to the RT simulation scenario, with the scene photograph and layout shown in Fig. \ref{Fig_Mea_scenario}. The measurements were conducted using a frequency-domain channel sounding platform  based on a vector network analyzer, spanning frequencies from 35-40 GHz with a total of 1001 frequency points. The Tx antenna was a biconical antenna with an approximately omnidirectional radiation pattern in the azimuthal plane, while the Rx antenna was a standard gain horn antenna (24.6 dBi gain, 10.67$^{\circ}$ horizontal HPBW). Both antennas were calibrated in an open-space environment, and their gain patterns were extracted. During the measurement, both Tx and Rx antennas were placed on mechanical turntables at a height of 1.4 m. The Tx antenna was moved horizontally in one-wavelength ($\lambda$) steps to form a sparse uniform linear array (ULA) with 50 elements, and for each Tx position, the Rx rotated in 10$^{\circ}$ increments for DSS.

\subsection{Measurement Results}

\begin{figure}
  \centering
  \subfloat[]{\includegraphics[width=4.3cm]{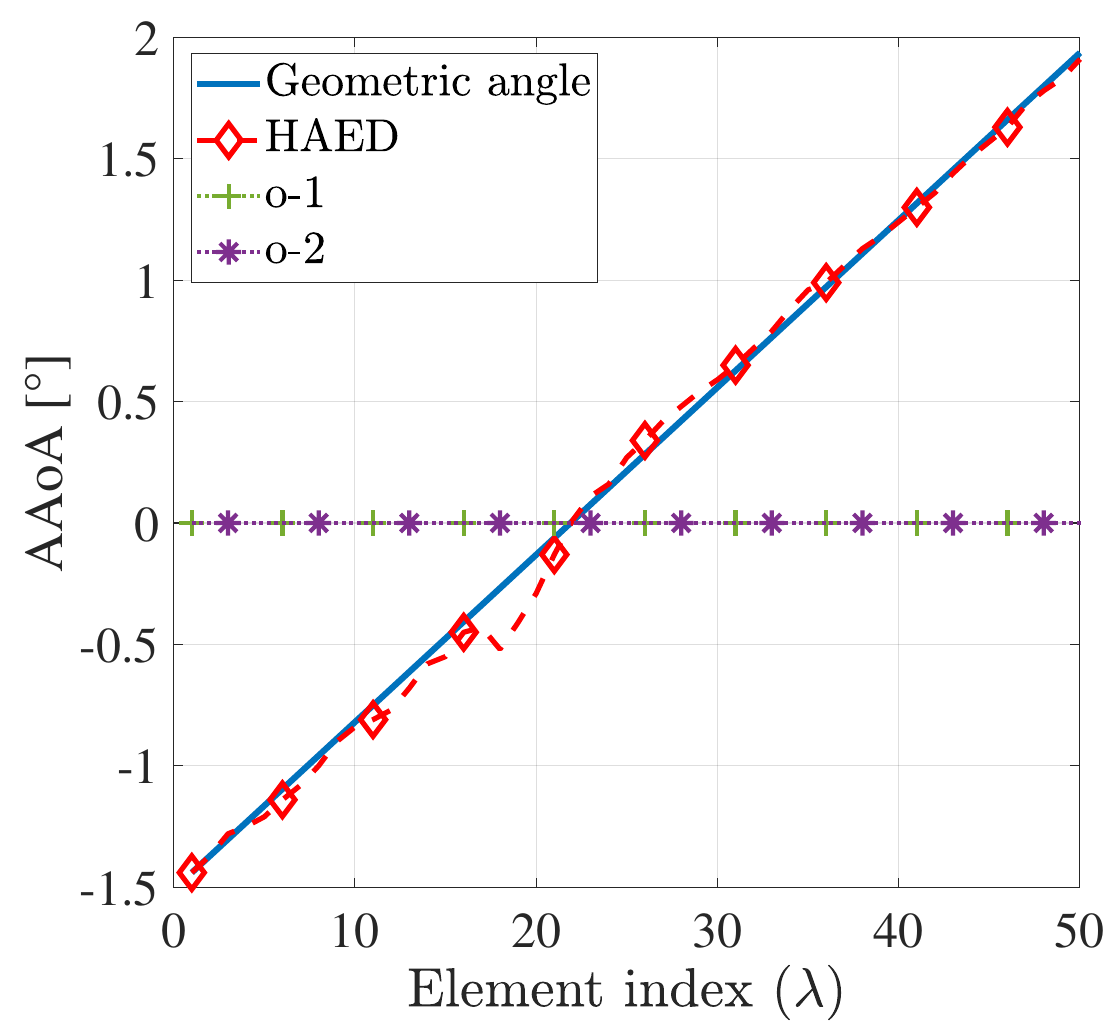} \label{Fig_Mea_MPC_a}}
  \hfill
  \subfloat[]{\includegraphics[width=4.3cm]{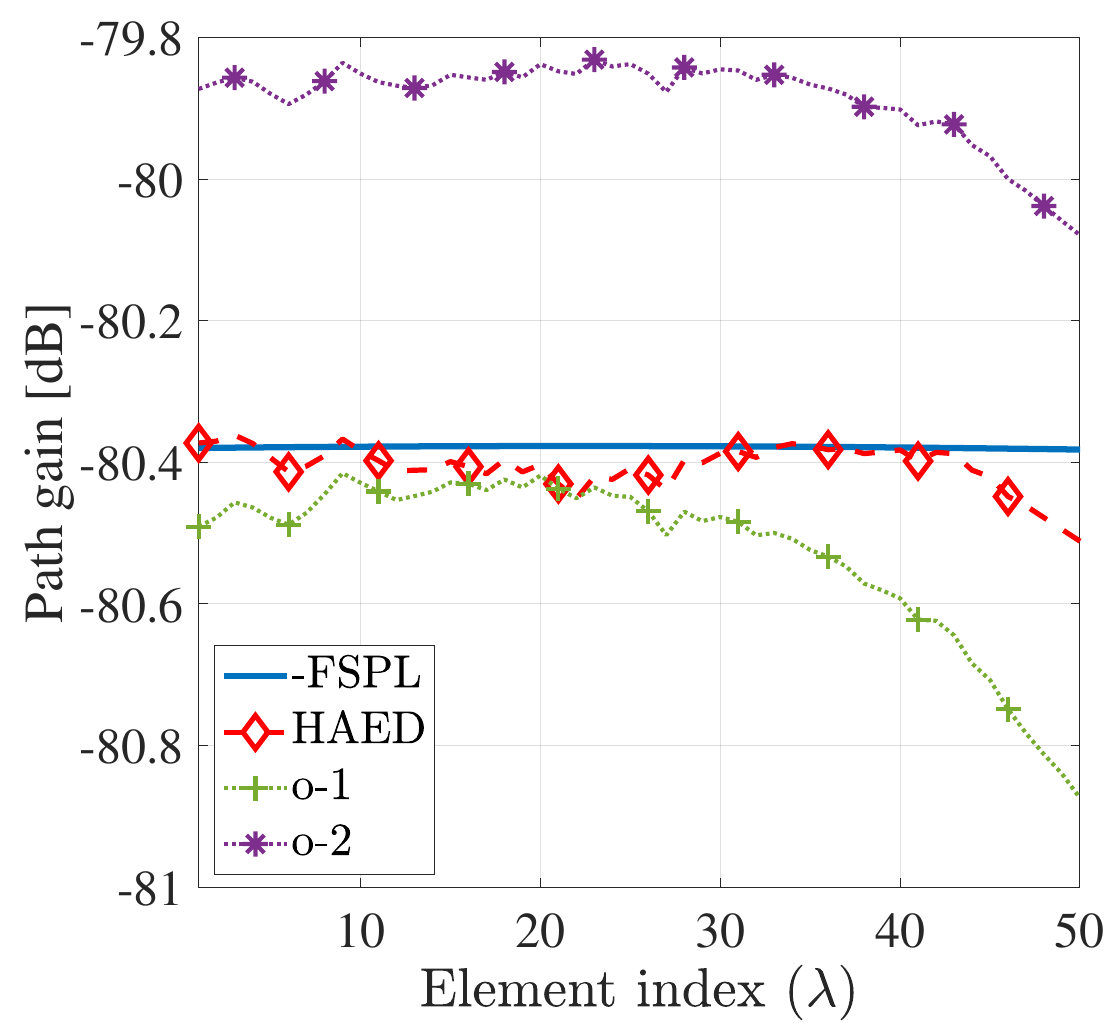} \label{Fig_Mea_MPC_b}}
  \\[-3.5mm]
  \subfloat[]{\includegraphics[width=4.3cm]{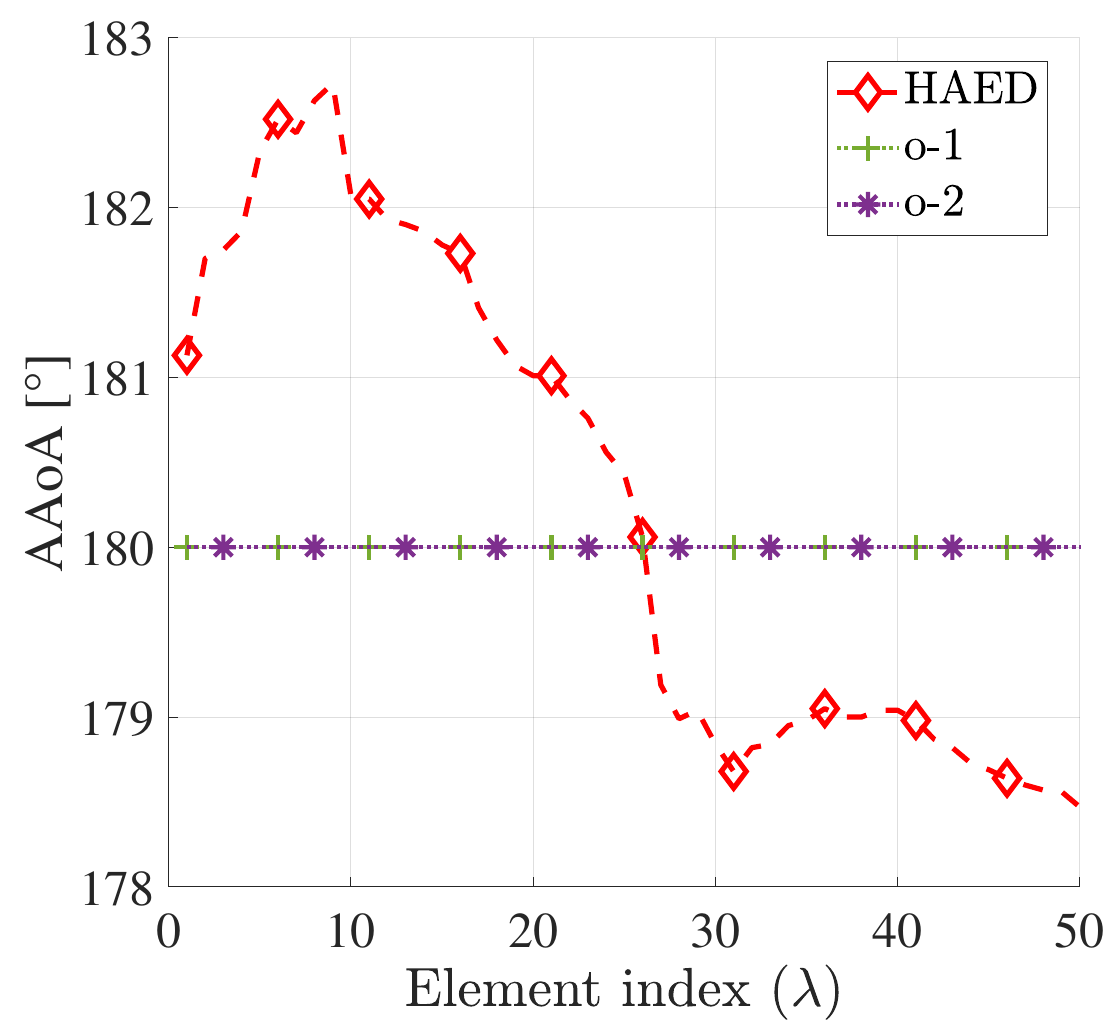} \label{Fig_Mea_MPC_c}}
  \hfill
  \subfloat[]{\includegraphics[width=4.3cm]{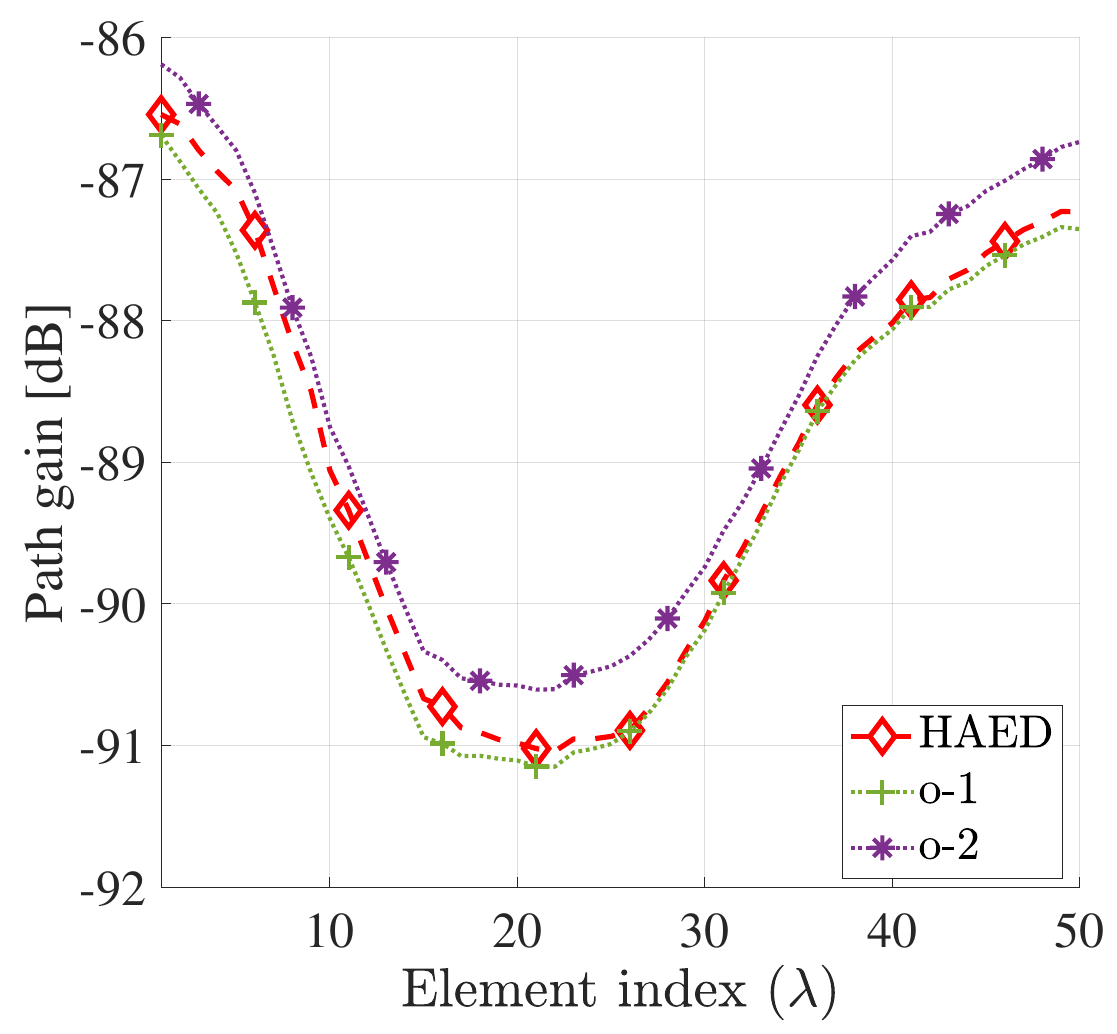} \label{Fig_Mea_MPC_d}}
  \caption{The Estimated AAoA and gain variation with Tx ULA elements. (a) AAoA and (b) gain of the LoS path (\#1). (c) AAoA and (d) gain of the NLoS path (\#2).}
  \label{Fig_Mea_MPC}
\end{figure}

To evaluate the accuracy of MPC angle estimation, we select a line-of-sight (LoS) path and a non-line-of-sight (NLoS) path, and analyze how the MPC's AAoA and power gains vary with the Tx ULA elements, as shown in Fig. \ref{Fig_Mea_MPC}. For the LoS path, the AAoA can be derived from the geometric positions of the Tx and Rx antennas, and the proposed HAED method closely matches the calculated geometric angle, as shown in Fig. \ref{Fig_Mea_MPC_a}. The average AAoA estimation error amounts to only $0.047^{\circ}$. It is observed that the AAoA variation is minimal, ranging from -1.5$^{\circ}$ to 2$^{\circ}$, yet the proposed method effectively captures these small changes, validating its outstanding performance in angle estimation. In contrast, traditional methods are limited by the spatial resolution of rotation and cannot capture the variation of AAoA across ULA elements, instead consistently estimating it at 0$^{\circ}$, which is physically unrealistic. Moreover, the LoS path gain estimated by the proposed HAED method lies closer to the negative free-space path loss (FSPL) curve, and the average power estimation error is just 0.03 dB. In comparison, gain estimated by the traditional o-1 method is consistently lower than that obtained via HAED, with a maximum difference of 0.33 dB. This discrepancy arises from an overestimation of directional antenna gain during the antenna decoupling. Conversely, the traditional o-2 method overestimates the LoS path gain with a maximum difference of 0.59 dB, due to the mismatch between the ASI and the HPBW (\( \text{ASI} < \text{HPBW} \) in this communication). Overall, the path gain calculated from the proposed HAED method always falls between the path gains obtained by the two traditional methods, as expected. For the NLoS path (\#2), the estimated AAoA varies continuously with the Tx ULA elements, and the path gain consistently lies between the values estimated by the two traditional methods, confirming the applicability of the HAED method to different types of MPCs.

\section{Conclusion}
\label{sec_5}
This communication proposes a high-resolution method for estimating MPC angle based on PADP for DSS, which leverages the power differences of MPCs received in adjacent scanning directions to precisely estimate the MPC's AAoA and amplitude or power. First, numerical simulations validate the performance of the proposed method. In the case of one MPC and two MPCs, the RMSEE of AAoA and amplitude estimated using the proposed HAED method approaches the corresponding $\sqrt{\rm{CRLB}}$, significantly outperforming traditional methods and improving the MPC angle and amplitude estimation accuracy by an order of magnitude. Additionally, extensive RT simulation results indicate that the proposed method estimates MPC's AAoA and power with near-perfect accuracy, reducing the AAoA estimation error by an average of 2.39° and the power estimation errors by 0.93 dB and 0.16 dB, respectively, compared to two traditional methods. Finally, channel measurement results at 37.5 GHz in an indoor scenario demonstrate that the proposed method can substantially improve MPC angle and amplitude estimation accuracy without increasing measurement complexity, achieving average AAoA and power estimation errors of only $0.047^{\circ}$ and 0.03 dB, respectively.

\bibliographystyle{IEEEtran}
\bibliography{paper}

\end{document}